# Fast, Accurate, Straightforward Extreme Quantiles of Compound Loss Distributions†

John Douglas (J.D.) Opdyke*

We present an easily implemented, fast, and accurate method for approximating extreme quantiles of compound loss distributions (frequency+severity) as are commonly used in insurance and operational risk capital models. The Interpolated Single Loss Approximation (ISLA) of Opdyke (2014) is based on the widely used Single Loss Approximation (SLA) of Degen (2010) and maintains two important advantages over its competitors: first, ISLA correctly accounts for a discontinuity in SLA that otherwise can systematically and notably bias the quantile (capital) approximation under conditions of both finite and infinite mean. Secondly, because it is based on a closed-form approximation, ISLA maintains the notable speed advantages of SLA over other methods requiring algorithmic looping (e.g. fast Fourier transform or Panjer recursion). Speed is important when simulating many quantile (capital) estimates, as is so often required in practice, and essential when simulations of simulations are needed (e.g. some power studies). The modified ISLA (MISLA) presented herein increases the range of application across the severity distributions most commonly used in these settings, and it is tested against extensive Monte Carlo simulation (one billion years' worth of losses) and the best competing method (the perturbative expansion (PE2) of Hernandez et al., 2014) using twelve heavy-tailed severity distributions, some of which are truncated. MISLA is shown to be comparable to PE2 in terms of both speed and accuracy, and it is arguably more straightforward to implement for the majority of Advanced Measurement Approaches (AMA) banks that are already using SLA (and failing to take into account its biasing discontinuity).

† The views expressed herein are those of the author and do not represent the views of GE Capital or any other institution.

* J.D. Opdyke is Head of Operational Risk Modeling at GE Capital. In addition to leading all operational risk modeling and quantification (capital estimation and stress testing) at GEC, he also leads model development for enterprise-level Economic Capital estimation, aggregation, and allocation across all risk types and lines of business (credit, operational, market, industrial, and insurance). J.D. previously has worked with some of the largest AMA banks globally during his 25 years as a quant, and his peer reviewed publications have earned multiple awards (including being voted Operational Risk Paper of the Year in 2012 and 2015). J.D. has earned undergraduate and graduate degrees from Yale and Harvard Universities, respectively, and has completed post-graduate work in robust statistics in the graduate mathematics department at MIT.

The author extends his sincere appreciation to Toyo Johnson, Nicole Opdyke, and Ryan Opdyke for their generous support and thoughtful insights.

## Background

Loss models for both insurance and operational risk capital estimation typically involve the estimation of extreme loss events – those that are very large, but associated with a very low probability of occurrence. One of the most common approaches is to estimate a compound or "aggregate loss distribution" (ALD) based on the convolution of two distributions: an independently estimated frequency distribution, representing the number of loss events that could occur within a given timeframe (typically a year), and an independently estimated severity distribution, representing the magnitude of these losses.[1] The quantile associated with a very large percentile of the ALD, such as the 99.9%tile per Basel II/III for operational risk regulatory capital, or the 99.5%tile for insurance per Solvency II, is the estimated extreme loss event. On average, these two examples represent a one-in-one thousand year loss and a one-in-two hundred year loss, respectively, and as such define the levels of capital financial institutions are required to hold per regulatory mandate; for economic and business planning, even higher percentiles typically are used.[2] Mathematically, this approach is defined below in (1):

$$\text{(annual) ALD} \sim S = \sum_{i=1}^{N} X_i \tag{1}$$

$$\text{where } X \sim f_X(x;\theta),\ N \sim p_N(n;\lambda) \sim \text{(typically) } \lambda^n e^{-\lambda}/n!, \text{ and } g_{X,N}(x,n) = f_X(x;\theta)\cdot p_N(n;\lambda)$$

$$\text{estimated capital} \sim VaR_\alpha(S) = \inf\left\{s \in \mathbb{R} : F_S(s) \geq \alpha\right\} \text{ for } \alpha \in (0,1), \text{ where } F_S(s) \text{ is the cdf of } S \ \ (\alpha = 0.999 \text{ for 99.9\%tile})$$

(i.e. the size of a loss, *X*, is distributed according to severity distribution *f* with parameter vector θ and the number of losses, *N*, is distributed according to frequency distribution *p*, which is typically the Poisson distribution with parameter *λ* representing the average number of losses in a year, and the two distributions are independent as the joint density equals their product; estimated capital is the quantile associated with the 99.9%tile of ALD, or *S*, which is the N-fold convolution of *X*).

In operational risk capital estimation under Basel II/III, use of the ALD is called "the loss distribution approach" (LDA) and is the most widely implemented of the AMA frameworks.[3] Unfortunately, extreme quantile estimation in this setting is complicated by the fact that the ALD rarely has a closed form, that is, it cannot be expressed exactly by a formula. Consequently, its specific quantiles must be approximated, and the gold standard for this approximation is extensive Monte Carlo simulation (see Opdyke and Cavallo, 2012a, 2012b). This involves simply simulating a random number of losses based on the estimated frequency distribution, and then randomly assigning values to each of those loss events based on the severity distribution.[4] This Monte Carlo simulation typically is "extensive" because many, many years' worth of losses must be simulated to adequately represent the extreme tail of the ALD, which is where the extreme quantiles (percentiles) that need to be estimated are located. For example, if an estimate of a one-in-one thousand year loss, on average, is required, an estimate based on only a thousand or even ten thousand years' worth of simulated losses will be inadequate: many millions of years must be simulated to adequately "fill in" the extreme tail of the ALD to obtain a reasonably precise approximation of that extreme quantile of that distribution.

---

[1] Independence between the frequency and severity distributions typically is assumed in these settings, even though it is widely acknowledged to be an unrealistic assumption (see Ergashev, 2008, and Chernobai et al., 2007). However, recent research from Stahl (2016) shows that when this assumption is violated, and a rigorous methodology properly accounts for and measures it, the magnitude of its impact on capital (quantile) estimation can be very large.

[2] The 99.97%tile is a typical value used for economic capital (almost all are 99.95%tile or above), based on a firm's credit rating, since it reflects 100% minus the historical likelihood of an AA rated firm defaulting within a one-year period (see Hull, 2012). Note also that, as shown in Opdyke (2014), the severity percentile associated with these ALD percentiles can be orders of magnitude larger — easily as high as the 99.999%tile and above. This holds enormous implications for the estimation of these percentiles, since as shown in Opdyke (2014), Shevchenko (2011), and herein, the variance of the estimate increases dramatically and nonlinearly in the size of the percentile (quantile) being estimated, which consequently can prevent its reliable estimation with extant data, regardless of the estimation method used.

[3] Virtually all AMA banks currently use the LDA for estimating their operational risk capital (see RMA, 2013). However, the Bank of International Settlements-Basel Committee on Banking Supervision (BIS-BCBS) recently has proposed abandoning the AMA for the Standardized Measurement Approach (SMA): a simpler, less risk sensitive framework (see BIS, 2016). The SMA has been strongly and universally criticized by operational risk measurement experts (see Mignola et al., 2016, and Peters et al., 2016, for just two examples), and its adoption has been postponed multiple times, arguably as a result of these critiques (see Osborn and Haritonova, 2017). Whether the framework that ultimately will be adopted includes the estimation of high quantiles of compound loss distributions remains to be seen, but the LDA approach is still widely used in many other areas of loss estimation, such as insurance and catastrophic loss models, and even credit risk models.

To be clear, the above refers only to "approximation error." When frequency and severity distributions are estimated based on samples of data, as is done in practice, there are three sources of error associated with the estimated extreme quantile of the ALD: modeling error, sampling error, and approximation error. Clearly understanding and distinguishing the distinct effects of each type of error on the ultimate quantile estimates is often overlooked, but absolutely necessary to the development of estimation and approximation methods that are reasonably accurate, reasonably precise, and reasonably robust.[5]

The first source of error – modeling error – is associated with estimating the right frequency and the right severity distributions, or in other words, selecting the right "model." For example, if the true severity distribution is a truncated Lognormal distribution, but a single sample of loss data generated from this distribution just happens to better fit a truncated LogGamma distribution (based on the best goodness-of-fit tests available[6]), the estimated extreme quantile will be wrong. The variance of this quantile estimate, over repeated samples, will reflect cases where the wrong distribution was selected, and so will be larger compared to the (unrealistic) case where we somehow "know" what the right distributions are. Unfortunately, it is widely recognized that statistical goodness-of-fit tests for selecting the "true" severity distributions have notoriously low statistical power in these settings. In other words, the "wrong" distribution is selected far more often than we would like due to several reasons: i) loss data is relatively scarce, especially for operational risk and many insurance and catastrophic loss settings, and so there is not enough of it to adequately discern between different distributions; and ii) loss data is often heterogeneous, that is, not generated from only one distribution, regardless of how the distribution is defined,[7] due to data quality issues and great challenges in dividing and classifying the data into homogenous "units of measure." Unfortunately, such "model error," which probably drives an enormous amount of the variance in the ultimate quantile estimate, largely has been ignored both by practitioners in the insurance and operational risk spaces, and in the relevant literatures. For a notable exception, see Mayorov et al. (2017).

---

[4] Again, this assumes independence between the frequency and severity distributions.

[5] The distinct effects of sampling and approximation error are shown empirically below in footnote 24.

[6] In this setting these tests typically are empirical distribution function-based (EDF-based) statistics, which are some function of the difference between the estimated cumulative distribution function (CDF) and the EDF. The most commonly used here are the Anderson-Darling (AD), the Cramér-von Mises (CvM), and the Kolmogorov-Smirnov (KS) tests.

[7] In other words, the challenge of heterogeneity remains regardless of the type of severity distributions which are used, be they single, fully parametric distributions; or body-tail split parametric distributions; or semiparametric distributions; or some function of fully nonparametric distributions, although the latter have been actively discouraged by regulators due to challenges in effectively extrapolating to extreme tails of the loss distribution when extant data samples are so limited.

A second source of error is the sampling error associated with estimating the parameters of the frequency and severity distributions. Even if the "true" frequency and severity distributions are assumed to have been selected, their parameters still must be estimated, and these estimates change from sample to sample because the data changes from sample to sample. Under textbook data assumptions (i.e. that data are independent and identically distributed, or "i.i.d."[8]), most estimators of these parameters will be unbiased, that is, centered on the true values, and be reasonably precise, but the estimates still will vary from sample to sample. However, even if it does not cause excessive variation in the estimated values of the severity parameters, this sampling variance, unfortunately, causes much larger variance in the estimates of the extreme quantiles of the severity distribution. This is due simply to the fact that, even under idealized data conditions (i.i.d. data), estimating extreme quantiles – especially of heavy-tailed distributions – requires excessive amounts of data to achieve any reasonable level of precision. This can be seen in the formula for the standard deviation of the quantile estimate in (2) below:

$$\text{standard deviation}\left(\widehat{Quantile_\alpha}\right) = \frac{\sqrt{\alpha\left(1-\alpha\right)}}{f\left(Quantile_\alpha\right)\sqrt{n}} \qquad (2)$$

where

$\alpha =$ specified percentile, a.k.a. 'confidence level' (here, $\alpha = 0.999$)

$Quantile_\alpha =$ quantile at specified percentile

$f\left(Quantile_\alpha\right) =$ value of probability density at the quantile of the specified percentile

$\left[\text{note that when the quantile itself is being estimated, this becomes } f\left(\widehat{Quantile_\alpha}\right)\right]$

$n =$ number of loss event data points in the sample

The standard deviation (the square root of the variance) is a function of the size of the quantile being estimated: the larger the quantile, the smaller the value of the probability density in the denominator of (2), and so the larger the variance of the quantile estimate. Similarly, we can obtain the number of loss events required (i.e. the sample size required) to achieve a specified level of precision in our quantile estimate, with (3) below.

$$n = \frac{4\alpha\left(1-\alpha\right)}{\varepsilon^2\left[f\left(Quantile_\alpha\right)\cdot Quantile_\alpha\right]^2} \qquad (3)$$

where

$\varepsilon = 2\cdot\text{standard deviation}\left(Quantile_\alpha\right)/Quantile_\alpha =$ relative error of the quantile estimate

$n =$ number of data points required to achieve precision of $\varepsilon\%$

Shevchenko (2011) uses (3) to provide a very conservative example, using a relatively light-tailed LogNormal severity distribution, under which 50,000 to 100,000 years' worth of loss data would be required to obtain estimates that are within 10% of the value of the true quantile (which arguably is not terribly precise). And of course, this assumes that "model error" does not exist, and that the loss data is i.i.d. – two very unrealistic assumptions. Violations of these strong assumptions unambiguously would increase the variance of the quantile estimate even more, and thus, increase the sample size requirements far beyond what are already unattainable levels of 50,000 to 100,000 years of losses to bring the precision of the quantile estimate again to reasonable levels. So it is generally gross understatement, by any standard, to say that the magnitude of sampling error associated with extreme quantile estimation is very large. In fact, it is not uncommon for the upper end of 95% confidence intervals on these estimates to be multiples of the size of the estimate itself! (see Opdyke, 2014, for specific examples, and a general treatment of estimation variance and bias in this setting).

---

[8] The most sweeping, yet common assumption in these settings is that the loss data is "i.i.d." – independent and identically distributed. "Independent" means that the values of losses are unrelated across time periods, and "identically distributed" means that losses are generated from the same data generating process, typically characterized as a single parametric statistical distribution. This is widely recognized as an unrealistic assumption made more for mathematical and statistical convenience than as a reflection of empirical reality. Some consequences of violations of this assumption are examined in Opdyke (2014).

Finally, in addition to model error and sampling error, the third source of error is "approximation error" which is strictly associated with the method of approximating the extreme quantile of the compound loss distribution. Put differently, this is the deviation that remains, when the approximation is compared to the true value of the extreme quantile, if we assume we have (always) selected the right severity and frequency distributions, and we have (always) perfectly estimated the right values of the parameters of these two distributions.

It is crucial to clearly identify each distinct source of estimation/approximation error here as a failure to do so prevents researchers from rightly identifying the causes of each, and from designing the most effective methods to address and minimize them.[9] While the materiality of modeling and sampling error cannot be understated in this setting, and is shown in other studies to be far larger than what most researchers have estimated due to Jensen's inequality (see Opdyke, 2014), this paper focuses only on approximation error.

So with our focus exclusively on approximation error, we return to approximating an extreme quantile of the ALD based on extensive Monte Carlo simulation, as described above. While this method is accurate in that it provides an unbiased approximation centered on the true value of the quantile, its precision is a function of the size of the quantile being estimated, and in this setting we are estimating very large quantiles.[10] Precision of the quantile approximation can be improved (i.e. the variance can be decreased) by increasing the number of simulations run, but the number of simulations required to achieve anything approaching a reasonable level of precision – for example, standard deviations that are not 50% or larger than the size of the quantile being approximated – is so large as to make this approach infeasible for most purposes in practice. However, to ensure the accuracy of very select results from other methods, it is still a very useful as a benchmark of "truth" when very large numbers of simulations are run (e.g. one billion year's worth), and that is how it is used in this paper.

[9] For example, with regards to approximation error in this setting, the discontinuity in Degen's (2010) widely used Single Loss Approximation (SLA), which under many conditions can materially and fatally bias the quantile (capital) approximation, was not reported in the literature until almost a half a decade later when Opdyke (2014) identified it and proposed a solution to fix it (that solution is generalized herein). Similarly, with regards to estimation error in this setting, even though the loss distribution approach had been in widespread use for many years, it was not until Opdyke and Cavallo (2012) that systematically upwards capital bias due to Jensen's inequality first was documented; and it was not until Opdyke (2014) that a proposed solution – the Reduced-bias Capital Estimator – made its way into the literature (see also Danesi et al., 2016). Without the proper identification and separation of very distinct sources of error, practical solutions will continue to elude researchers and practitioners for unnecessarily long periods of time.

[10] As mentioned above, while the percentiles being estimated for the ALD are very large at, for example, the 99.9%tile, this corresponds to estimating even higher percentiles of the severity distribution, such as 99.999%tile and larger. The point here is that the challenges of obtaining reasonable precision in this estimation are even greater than they appear at first glance.

In assessing alternate methods, we consider, in addition to unbiasedness (accuracy) and precision, both speed and ease of implementation, keeping in mind current widespread practice when evaluating the latter. Our goal in this paper is to develop a method for approximating extreme quantiles of the ALD that is not only most accurate and precise under the widest range of conditions, but also the fastest (i.e. most computationally efficient), the most straightforward to use, implement, and understand, *and* the most consistent with widespread current practice in the relevant industries. This final criterion is important as even the "best" methods by other criteria will not be widely used if they are too difficult to understand and/or implement vis-à-vis how much they disrupt the current implementations of most institutions and practitioners.

## Methods of Approximating Extreme Quantiles of Compound Loss Distributions

Below are some of the published methods for approximating extreme quantiles of the ALD:

- Extensive Monte Carlo simulation (see Opdyke and Cavallo, 2012a, 2012b)
- Fast Fourier Transform (FFT, see Embrechts and Frei, 2009)
- Panjer Recursion (see Panjer, 2006)
- Direct Method (see Kato, 2012)
- Wavelet Expansion Methods (see Ishitani and Sato, 2013)
- Single Loss Approximations (see Böcker and Klüppelberg, 2005; Böcker and Sprittula 2006; Sahay et al., 2007; Degen, 2010; and Peters et al., 2013)
- Interpolated Single Loss Approximation (ISLA) (see Opdyke, 2014)
- Perturbative Expansion Approach (Hernandez et al., 2014)

The criteria by which we examine and/or evaluate these methods include 1. accuracy (i.e. unbiasedness and/or consistency), 2. precision (i.e. statistical efficiency), 3. speed (real runtime, as well as cpu runtime if systematically divergent from real runtime), and 4. implementation considerations: is the method straightforward to use, understand, and consistent with widespread current practice in the relevant industries?

As described above, extensive Monte Carlo simulation is very straightforward to implement, understand, and verify. It is considered a gold standard vis-à-vis accuracy because it is unbiased, and it also provides arbitrary closeness to true quantile values. However, its precision is only reasonably small if very large numbers of simulations are run, making this method prohibitively slow in almost all practical settings.

The Fast Fourier Transform (FFT) and Panjer recursion both have been widely used in this setting, and both are known to be unbiased. To achieve the same level of precision, FFT is widely cited as being much faster and more computationally efficient than Panjer recursion (see Embrechts and Frei, 2009), but both methods require algorithmic looping and are slower than the non-looping, SLA-based analytic approximations examined below.[11] In addition, as noted in de Jongh et al. (2016), neither of these methods are as easily implemented as the SLA-based formulaic approximations as their responsible use requires a priori knowledge of appropriate values of input parameters.

[11] The speed of these two methods is not explicitly tested in this paper, but as they require algorithmic looping, they cannot be as fast as closed-form approximations, and have been shown to be slower in a number of publications that track computer runtimes (see, for example, Gollini and Rougier, 2016, and Temnov and Warnung, 2008. And this also is noted in de Jongh et al., 2016).

Neither the Direct Method of Kato (2012) nor the Wavelet Expansion methods of Ishitani and Sato (2013) are widely used in the relevant insurance and operational risk settings. They are not explicitly implemented herein due to their complexity, but based on their reliance on numeric integration and algorithmic looping, both are much slower than the SLA-based methods examined below (see their runtimes listed in both publications). Also, unlike the SLA-based approximations, their runtimes are an increasing function of the sample size, which runs counter to a need for larger samples to increase statistical precision. So we suggest that if the accuracy of the widely used and easily implemented SLA-based methods is beyond need for improvement, then there arguably would be no need to turn to these slower, more complex methods.

Finally, the SLA-based analytic approximation methods that include the Single Loss Approximation (SLA) method of Degen (2010), the Interpolated Single Loss Approximation method (ISLA) of Opdyke (2014), and the modified ISLA approach developed herein (MISLA) are examined in detail below. As the most promising competitor to these approaches, the perturbative expansion of Hernandez et al. (2014) also is implemented and benchmarked for speed and accuracy against all three SLA-based methods.

## The Single Loss Approximation Approaches: SLA, ISLA, and Modified ISLA (MISLA)

### SLA

Böcker and Klüppelberg (2005) were the first to propose a Single Loss Approximation approach to approximating the extreme quantile of the ALD. This was followed by Böcker and Sprittula (2006) and a similar but more specific derivation by Sahay et al. (2007) based on the Generalized Pareto distribution tail. The results of the first two papers were based on empirical analyses, but Degen (2010) eventually subjected the approach to general and rigorous analytic derivations. So it is his version of SLA, expanded to include conditions of infinite mean, that is used herein and presented below (these all are shown graphically in Figure 1):

$$\text{if } \xi < 1, \qquad C_\alpha \approx F^{-1}\left(1-\frac{1-\alpha}{\lambda}\right)+\lambda\mu \tag{4.a}$$

$$\text{if } \xi = 1, \qquad C_\alpha \approx F^{-1}\left(1-\frac{1-\alpha}{\lambda}\right)+\lambda\mu_F\left[F^{-1}\left(1-\frac{1-\alpha}{\lambda}\right)\right] \tag{4.b}$$

$$\text{if } 1 < \xi < 2, \quad C_\alpha \approx F^{-1}\left(1-\frac{1-\alpha}{\lambda}\right)-\left(1-\alpha\right)F^{-1}\left(1-\frac{1-\alpha}{\lambda}\right)\cdot\left(\frac{c_\xi}{1-1/\xi}\right) \tag{4.c}$$

where

$c_\xi = \left(1-\xi\right)\dfrac{\Gamma^2\left(1-1/\xi\right)}{2\Gamma\left(1-2/\xi\right)}$ ; $\mu_F\left(x\right)=\int_0^x\left[1-F\left(s\right)\right]ds$ ;[12] $C_\alpha$ = "capital" ;

$\alpha =$ "confidence level" (e.g. $\alpha = 0.999$ for Regulatory Capital)

[12] Note that Simpson's method is very straightforward and works very well to carry out the numeric integration required to calculate $\mu_F\left(x\right)$ . This is due to the fact that the functions being integrated – parametric cumulative distribution functions – are very well behaved, i.e., very smooth, which is ideally suited to Simpson's method.

$F^{-1}(\cdot)$ is the quantile function of the severity distribution ; $\lambda$ is the (typically Poisson) frequency parameter ; $\mu$= mean of severity distribution ; $\xi$= the tail index; [13] and $\Gamma(\cdot)$ is the gamma function.

Note that $\xi \geq 2$ is so extreme as to be irrelevant in these settings, and that the Poisson frequency distribution is assumed in the above presentation of the SLA. Only minor changes are required to adapt SLA for other frequency distributions, but the Poisson is the widespread, acknowledged default in this setting. A number of publications have shown that the choice of frequency distribution, as well as the parameter estimation of the frequency distribution, has relatively little effect on quantile estimation for compound loss distributions (see Opdyke, 2014). In other words, the selection and parameter estimation of the severity distribution accounts for almost all of the value of the ultimate quantile estimate of the ALD.

SLA is extremely fast to implement because it is a closed-form approximation, i.e. a formula rather than an algorithm, and it has been tested in the literature and shown to be sufficiently accurate under many conditions (see Opdyke, 2014, and Hess, 2011). Unfortunately, as described in Opdyke (2014), a notable flaw mars its otherwise very useful implementation: as the tail index of the severity distribution approaches a value of one from above or below, the quantile (capital) approximation diverges to negative or positive infinity, respectively (see Figure 1 below). Importantly, this divergence is not an asymptotic result: it is not affected by sample size in any way. The divergence from below is due to the fact that the severity mean ($\mu$) in the second term of (4.a), otherwise known as the "correction term," becomes infinitely large, by definition, as the tail index approaches one; and divergence from above is due to the fact that the denominator in the last factor of the correction term of (4.c), i.e. $1-1/\xi$ , becomes infinitely small, making the entire correction term, when subtracted, infinitely negatively large. So we have divergence to positive and negative infinity in the quantile/capital estimate as the tail index approaches a value of one from below and above, respectively. The closer the estimated tail index is to one, the worse is the divergence, and the more biased is the quantile/capital estimate.[14] This bias is unbounded, as tail index estimates can be arbitrarily close to a value of one, and so quantile/capital estimates under SLA can and often do diverge very materially and systematically from their true values. For example, for the Generalized Pareto Distribution with parameter estimates of θ=10,000 and $\xi$ =0.995, and thus, a tail index= 0.995, the SLA-based quantile/capital estimate for VaR99.9 is $288,851,090, when the true VaR99.9 is approximately $240,825,166. Worse still, for the LogGamma distribution with parameter estimates of α=4 and β=1.0050251256=1/0.995, and thus, the same tail index=0.995, the SLA-based quantile/capital estimate for VaR99.9 is $40,022,601,637, when the true VaR99.9 is approximately $22,690,333– a difference of more than 3 orders of magnitude.[15] While such estimates arguably are not commonly encountered in practice, they also are not extremely rare. Clearly, the unbounded nature of the systematic bias in SLA-based quantile/capital estimates can lead to unacceptably large inaccuracies in these settings.

[13] Tail indices for all severity distributions examined in this paper are defined in the Appendix. Degen (2010) defines the tail index as follows: A positive measurable function $f$ is regularly varying with parameter $\beta$ (written as $f \in RV_{\beta}$ ) if $f$ satisfies $\lim_{t\to\infty} f(tx)/f(t) = x^{\beta}$ for all $x > 0$ . In the case of a probability density $f$ , one in particular has that $f \in RV_{-1/\xi-1}$ implies $1-F \in RV_{-1/\xi}$, where $\xi$ is the tail index.

[14] Note that reliance on higher order terms do not "fix" this discontinuity in Degen's (2010) SLA.

[15] Both "true" VaR99.9 values here are based on Monte Carlo simulations of one billion years of losses with λ=25.

Of course, this does not apply to severity distributions that cannot have infinite mean, such as the LogNormal. An SLA implementation based on one of these severity distributions utilizes only (4.a), and because the mean of the distribution ($\mu$) cannot diverge to infinity, SLA does not diverge to infinity. However, it is important to note that for all severity distributions that _can_ have infinite mean (e.g. the Generalized Pareto distribution (GPD), the LogGamma distribution, and many others), even when their parameter estimates are restricted to values that prevent an infinite mean, the SLA discontinuity remains a major and relevant flaw. It is easy to see why this is the case by examining the first frame of Figure 1. Take the case of a GPD where its estimated tail index parameter, $\hat{\xi}$, is restricted to values of less than one. The parameter still can take on values that are arbitrarily close to one, and thus, provide quantile (capital) estimates that diverge to positive infinity. In fact, this capital bias could happen even more often under constrained parameter estimations because estimates of $\xi$ that would be larger than one, and often farther from 'the bias zone' around one, actually will be forced to be arbitrarily close to one by the constrained estimation, and thus, produce highly biased quantile (capital) estimates.

The divergence of SLA described above is straightforward and easily verified. One reason it was not noted earlier[16] could be that the size of the divergence is not large unless tail indices are close to one. However, not only does this remain a problem analytically even under asymptotic conditions, but also when conducting simulations to generate thousands of capital estimates, as is very commonly done in practice. All it takes is several, or even just one random sample of losses to generate an estimated tail index parameter, $\hat{\xi}$, arbitrarily close to one (which is very likely to occur at least once in thousands of simulations), for descriptive statistics of the entire set of simulations, such as its mean and its standard deviation, to be notably biased and NOT reflective of the true capital (quantile) distribution. An example of this is shown in a simple simulation study in the Results section below.

So bias due to the discontinuity of Degen (2010) when $\xi$ approaches values of "one," from above or below, is A. non-asymptotic,[17] B. relevant under conditions of both finite and infinite mean, and C. even relevant when the estimation of severity parameters of distributions that _can_ have infinite mean are restricted to values that only generate finite means. The only conditions under which this divergence is not relevant occur when only (4.a) is used, that is, when the entire pool of severity distributions excludes ALL that can have infinite means. Given the heavy-tailed nature of compound loss distributions in the operational risk, insurance, and catastrophic loss settings, this is rarely, if ever, the case, and likely would be rejected outright under regulatory review of an AMA-LDA (or similar) quantification framework.

---

[16] Opdyke (2014) is the first, and only, paper to note the discontinuity in SLA and explain the consequent systematic bias in SLA-based capital (quantile) estimates. Even otherwise thorough and recent treatments of extreme quantile approximation techniques for compound loss distributions (see de Jongh et al., 2016) fail to explain _why_ SLA falls short, under many conditions, as an approximation method in this setting. Absent an understanding of why a method underperforms, it is difficult, if not impossible to know the exact conditions that will cause its underperformance, let alone know how to fix it, or to be aware of why, how, and exactly where its competitors will outperform it.

[17] Note that this is non-asymptotic as it relates to sample sizes, as $n \to \infty$. But the derivation of Degen (2010) absolutely is an asymptotic result vis-à-vis the size of the percentile/quantile being estimated, i.e. valid as $\left(100\% - \alpha\right) \to 0$.

**Figure 1: SLA, with ISLA Interpolation for SLA Divergence at $\xi$ = 1 for GPD Severity (θ = 50,000)**

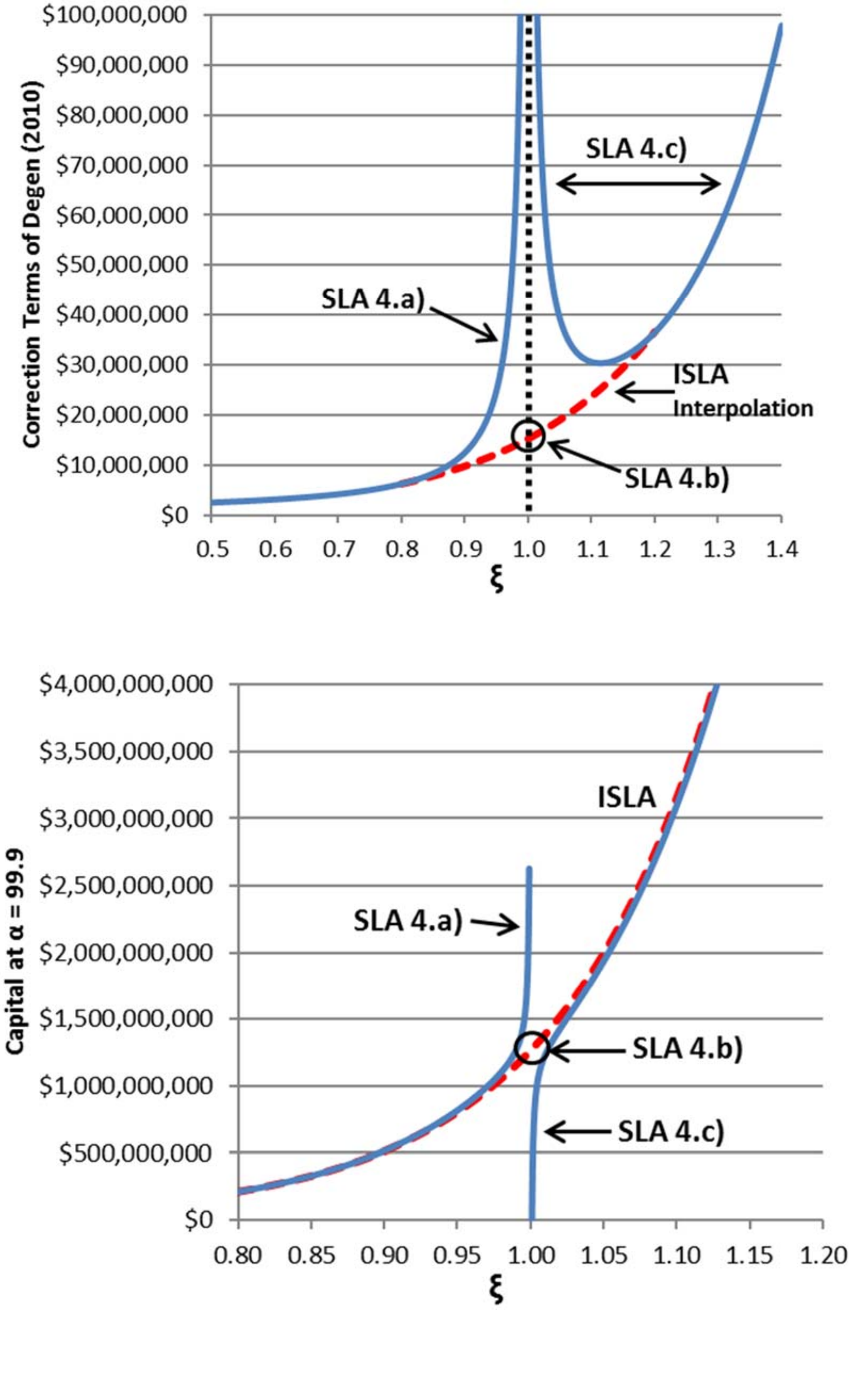

## ISLA and MISLA

Yet all is not lost. One part of Degen's (2010) definition of SLA may save it by providing an effective mechanism to avoid the divergence-induced capital (quantile) bias. When the tail index exactly equals a value of one, the quantile equals (4.b), as given. Although it is highly unlikely in practice, and theoretically impossible that any tail index estimate will exactly equal 1.0 when modern computer chip precision allows estimation to sixteen decimal places, the theoretical result is important because it provides an anchor by which one can reliably interpolate from values below one to values above one, i.e. as a bridge from (4.a) to (4.c). This would allow us to effectively cross over the "bias zone" near one, and provide an approximation of quantile values within this zone. So if (4.a) is used as $\xi$ approaches one from below, and (4.c) is used when $\xi$ approaches one from above, and (nonlinear) interpolation is used in between by connecting both to (4.b) when $\xi$ gets sufficiently close to the bias zone, then we avoid the discontinuity and the capital (quantile) bias it causes. And this is very similar to what Opdyke (2014) uses to avoid the bias caused by SLA divergence. Opdyke (2014) uses a nonlinear interpolation from (4.a) to (4.c), and determines the endpoints of each, that is, where $\xi$ gets sufficiently close to one on either side, empirically, by simply generating the graph shown in Figure 1. This differs from the current proposal – Modified ISLA (MISLA) – only in that MILSA explicitly uses (4.b) to anchor the mid-point of the interpolation, so that one interpolation is performed from below $\xi = 1.0$, from (4.a) to (4.b), and one is performed from above $\xi = 1.0$, from (4.b) to (4.c). This is defined in (5) below. For most severity distributions, there is virtually no difference between ISLA and MISLA empirically, as shown in the results section below. But for one severity distribution identified thus far – the LogGamma – the difference in the interpolation can be noticeable, as shown in Figure 2 below. Hence the need to generalize ISLA with MISLA by implementing two nonlinear interpolations rather than just one. Note, however, in Figure 2 that the difference in the total capital estimates of ISLA vs. MISLA is barely noticeable as total capital is orders of magnitude larger than the correction term. Although this is typically the case, MISLA remains the safer and more accurate approximation.

MISLA is defined below: (5)

For severity distributions that cannot have infinite mean: Quantile Estimate = (4.a) of Degen (2010).

For severity distributions that can have infinite mean (whether or not the parameter estimation is restricted to conditions of finite mean):

where $\hat{\xi}$ is the tail index based on a severity parameter estimate from the data sample,

**(5.a)** when, for $\hat{\xi}_{low}$ =low interpolation point, $\hat{\xi} \leq \xi_{low} < 1$: Quantile Estimate = (4.a) of Degen (2010).

**(5.b)** when, for $\hat{\xi}_{high}$ =high interpolation point, $1 < \xi_{high} \leq \hat{\xi}$ : Quantile Estimate = (4.c) of Degen (2010).

**(5.c)** when $\xi_{low} < \hat{\xi} < 1$: Quantile Estimate = $C_{\alpha} \approx F^{-1}\left(1 - \frac{1-\alpha}{\lambda}\right) + ICT_{low}$ where

Interpolated Correction Term = $\mathrm{ICT}_{low} = \left[ \left(\mathrm{LCT}\right)^{(1/\mathrm{Root})} + \mathrm{CIR} \times \mathrm{DRS} \right]^{\mathrm{Root}}$

Difference Root Scale = $\text{DRS} = \dfrac{(\text{HCT})^{(1/\text{Root})} - (\text{LCT})^{(1/\text{Root})}}{\text{FRC}}$

High Correction Term = $\text{HCT} = \lambda\mu_F \left[ F^{-1}\left(1 - \dfrac{1-\alpha}{\lambda}\right)\right]$

Low Correction Term = LCT = $\lambda\mu_{low}$ where $\mu_{low} = E\left[\text{severity distribution with parameter values } \to \xi_{low}\right]$

Count in Range = $\text{CIR} = \left(\hat{\xi} - \xi_{low}\right)\cdot \text{PRE}$ , Full Range Count = $\text{FRC} = \left(1 - \xi_{low}\right)\cdot \text{PRE}$ , PRE = 1000, Root=50

**(5.d)** when $1 < \hat{\xi} < \xi_{high}$ : Quantile Estimate = $C_\alpha \approx F^{-1}\left(1 - \dfrac{1-\alpha}{\lambda}\right) + ICT_{high}$ where

Interpolated Correction Term = $\text{ICT}_{high} = \left[(\text{LCT})^{(1/\text{Root})} + \text{CIR}\times\text{DRS}\right]^{\text{Root}}$

Difference Root Scale = $\text{DRS} = \dfrac{(\text{HCT})^{(1/\text{Root})} - (\text{LCT})^{(1/\text{Root})}}{\text{FRC}}$

High Correction Term = $\text{HCT} = (1-\alpha) F^{-1}\left(\xi_{high}; 1 - \dfrac{1-\alpha}{\lambda}\right)\cdot\left(\dfrac{c_{\xi_{high}}}{1 - 1/\xi_{high}}\right)$ where $c_{\xi_{high}} = \left(1 - \xi_{high}\right)\dfrac{\Gamma^2\left(1 - 1/\xi_{high}\right)}{2\Gamma\left(1 - 2/\xi_{high}\right)}$

Low Correction Term = $\text{LCT} = \lambda\mu_F \left[ F^{-1}\left(1 - \dfrac{1-\alpha}{\lambda}\right)\right]$

Count in Range = $\text{CIR} = \left(\hat{\xi} - 1\right)\cdot \text{PRE}$ , Full Range Count = $\text{FRC} = \left(\xi_{high} - 1\right)\cdot \text{PRE}$ , PRE = 1000, Root=50

Empirically, PRE=1000 gives sufficient precision in the interpolation, and Root=50 gives sufficient curvature. Attempts to numerically optimize these values add unnecessary complexity, which increases runtime with essentially no gain in accuracy. In other words, numeric optimization of these parameters would not maximize the objective function for our selection of an interpolation method in this setting: use the most accurate interpolation, to the point of diminishing returns, that does not materially increase the speed or complexity of implementation (which remains consistent with the approximation framework that, by far, is most widely used in practice). On the other extreme, it is true that in many cases, just using a linear interpolation will yield ultimate quantile/capital approximations very close to the true values. But for some severities and some (extreme) quantiles, this over-simplicity comes at the price of material inaccuracy, so the insurance of using the nonlinear interpolation of MISLA is worth the negligible tradeoff in complexity. In the end, the nonlinear interpolation of MISLA defined above maximizes our objective function while ensuring the highest levels of accuracy: it remains straightforward, fast, and consistent with widespread practice, without succumbing to the temptation to oversimplify the approach (e.g. linear interpolation) and in the process, sometimes materially sacrificing the

**Figure 2: SLA, with MISLA and ISLA Interpolations at $\xi$ = 1 for LogGamma Severity (α = 4.89)**

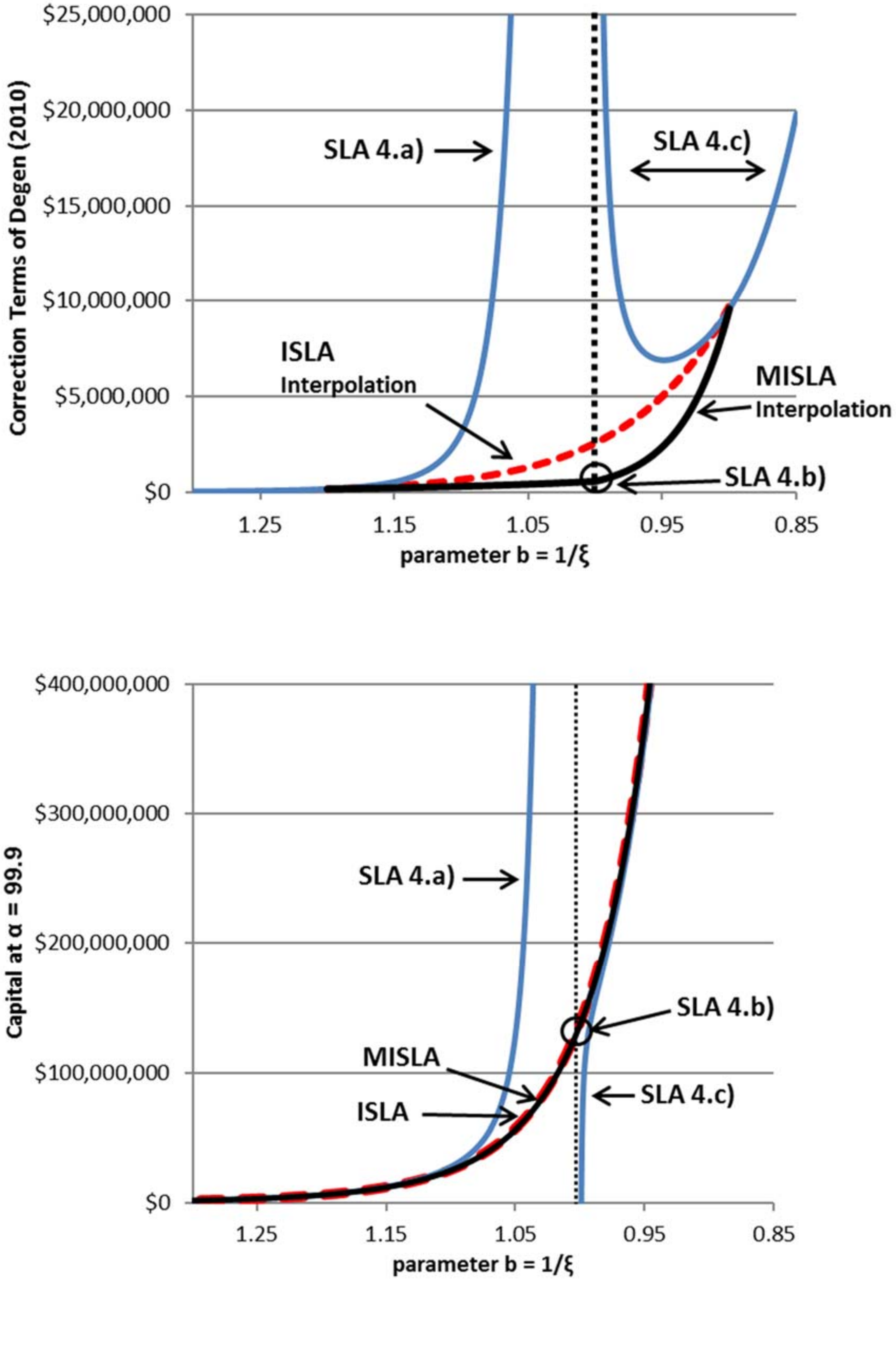

accuracy of the approximation.[18]

The Perturbation Approach

Hernandez et al. (2014) took an entirely different approach compared to SLA-based methods. They truncate a perturbative expansion ("PE") of the percentile of the aggregate loss distribution at different orders, although we found that orders beyond the second make negligible difference empirically in the approximation (hence, "PE2"). Terms of the expansion include lower partial moments (LPMs), which often require numeric integration. However, with upper bounds, this integration generally is much faster than most of the numeric integration required by the other methods listed above. And when implementing PE2 in the Results section below, wherever possible, depending on the severity distribution used, we employ for these LPMs analytic solutions or algorithms faster than numeric integration (for example, in some cases, continuous fractions)[19] to ensure its fastest possible execution. Also, in de Jongh et al. (2016) PE2 was favorably reviewed in terms of its speed and accuracy against widely used alternatives (e.g. FFT). So for all of these reasons, we implement it below as the best potential competitor to MISLA. Results for ISLA and SLA also are shown below.

## Empirical Study

We compare the quantile approximation methods using aggregate loss distributions (ALDs) with a Poisson frequency distribution ($\lambda = 25$) and twelve different severity distributions, several of which are truncated.[20] These include the Generalized Pareto, LogGamma, LogNormal, LogLogistic, Frechet, BetaPrime, Inverse Gamma, Paralogistic, and the Inverse Paralogistic distributions. Truncated versions of the first three, which are commonly used in operational risk capital modeling, also are used, with data collection thresholds of 5,000 (details of all twelve distributions, including their tail indices, $\xi$ , are shown below in the Appendix). The Frechet, BetaPrime, Inverse Gamma, Paralogistic, and the Inverse Paralogistic distributions are less commonly used in these settings compared to the other seven,[21] but as two-parameter distributions, they remain viable candidates in light of recent regulatory guidance (see OCC, June, 2014) that emphasizes the benefits of using severity distributions with fewer parameters, ceteris paribus, to avoid overfitting (this is in contrast to the use of, say, the Champernowne (see Bolancé et al., 2012), g-and-h (see Dutta and Perry, 2006), Log Normal Inverse Gaussian (see de Jongh et al., 2016), and Generalized Beta of the Second Kind (see Dutta and Perry, 2006) distributions, which have 3, 3, 4, and 4 parameters, respectively). However, the main point here is to demonstrate the applicability of MISLA across a wide range of relevant severity distributions, regardless of the exact setting in which very high quantiles must be estimated, be it operational risk modeling, insurance modeling, catastrophic loss modeling, or even credit risk modeling. ISLA (Opdyke, 2014) only was tested on three severity distributions, so we are subjecting his general interpolation approach (with improvements) to a much more thorough empirical testing process, while also comparing it to a completely different, yet well-reviewed approach (i.e. the PE2).

---

[18] It should be noted that (4.b) from Degen (2010) could be used even when $\xi < 1$ to obtain essentially identical results to MISLA. That is, one could use (4.b) when $\hat{\xi} \leq 1$ , and interpolate from (4.b) to (4.c) when $\hat{\xi} > 1$ instead of interpolating both above and below $\hat{\xi} = 1$ as is required of MISLA. However, MISLA preserves the use of (4.a) from Degen (2010), which is consistent with earlier, widely circulated derivations of SLA (see Böcker and Klüppelberg (2005) and Böcker and Sprittula (2006)), is most widely understood in the industry, and is most widely used in practice. So using two interpolations instead of one is more reflective of actual implementations in widespread practice. And MISLA's formula is consistent across all potential values of $\hat{\xi}$ , differing only in its correction term, ICT.

[19] For example, for the LogNormal and Truncated LogNormal distributions, the Erf function was used to calculate the LPMs, and the CDF function of the Generalized Pareto and Truncated Generalized Pareto distributions was used to calculate the respective LPMs of each. For the Frechet and Inverse Gamma distributions, a continuous fraction, rather than numeric integration, was used to calculate values for the upper incomplete gamma function.

[20] Note that regulators in the U.S., in addressing the issue of data collection thresholds, have rejected so-called shifted distributions in favor of truncated distributions (see Opdyke, 2014, Danesi et al., 2016, and Ergashev et al., 2016).

Note that of all the distributions listed above, only the LogNormal and Truncated LogNormal cannot have infinite means, but results for these two distributions also are included to demonstrate the performance of SLA (which is identical to ISLA and MISLA, by definition, under these conditions) and allow for its comparison to PE2.

Approximations of two different percentiles of the ALD are used: the 99.9%tile, which corresponds to operational risk regulatory capital under an AMA framework, and the 99.97%tile, which is commonly used as the estimate of Economic Capital (see Opdyke, 2014).

Because speed is an important criterion of comparison of MISLA and PE2, we use the statistical software package of the largest privately owned software firm in the world – SAS® – to implement both methods.[22] For all results, extensive Monte Carlo simulation of one billion simulated years of loss events is used as a comparison benchmark of "truth." Complete results are shown in the Results section below.

Parameter values for all distributions are listed below in Table 1, and these were chosen to in most cases yield roughly the same mean across distributions, when possible and reasonable, but always to yield exactly the same tail index value of $\xi = 0.99$. This is sufficiently close to a value of one to observe notable divergence in SLA from the true quantile (capital) value. Of course, when based on samples of loss event data, estimates of $\xi$ (which always rely directly on parameter estimates of the severity distribution, as shown in the Appendix), can be arbitrarily close to $\xi = 1.0$, and thus exhibit far more bias than is shown to be associated with SLA when $\xi = 0.99$. Results for values of $\xi$ above one, such as $\xi = 1.01$, were very similar in absolute magnitude (but with the opposite sign) to those associated with $\xi = 0.99$, and consequently are not presented herein. Also, we chose to present only $\xi = 0.99$ because some financial institutions choose to include "infinite mean" severity distributions in their severity selection pools only when constraining their parameter estimation to always yield $\hat{\xi} < 1.0$. So the bias inherent in the use of SLA in such a framework will be demonstrated effectively by using cases of $\xi = 0.99$.

[21] However, with the exception only of the BetaPrime distribution, these all are cited in major texts on the topic (see Loss Models: Further Topics, Klugman et al., 2013, and Loss Models: From Data to Decisions, Klugman et al., 2012).

[22] SAS and all other SAS Institute Inc. product or service names are registered trademarks or trademarks of SAS Institute Inc. in the USA and other countries.

Finally, as mentioned above, it is important to note that our understanding of why SLA falls short as an approximation method – i.e. due to its discontinuity – is critical here as it allows us to focus our study much more efficiently on the right conditions (i.e. $\xi = 0.99$ , the 'bias zone') than if we did not have this knowledge. Absent this understanding, we would have to cover far more extensive and disparate loss data conditions, many of which would be redundant and unnecessary. And even at the end of implementing such an exhaustive empirical study, we would have no further understanding of the mechanics underlying SLA's imperfect performance: only a less-than-perfect awareness of where it did and did not perform well, which could leave lingering questions as to whether we had covered a broad enough range of data and distributional possibilities relevant to the specified settings and the analytical problem being solved. So understanding the analytics behind SLA, as they are shown simply in Figure 1, is of crucial importance here.

For MISLA and ISLA, the interpolation endpoints for each severity distribution were chosen based on visual inspection of the graph of (4.a) and (4.c) shown for versions of Figure 1 generated for each severity distribution: the discontinuity of the SLA-based quantile approximations from (4.a) and (4.c) begin around values of $\xi = 0.85$ from below and $\xi = 1.15$ from above, respectively, for most of the severities examined. Graphing (4.a) and (4.c) for several of the severity distributions show minor deviations from these two values, typically within ±0.05. Since the typical size of a severity distribution pool is only half a dozen distributions or less, graphing (4.a) and (4.c) for each one to determine these endpoints is not an onerous task. But we also include results herein for the case where the interpolation endpoints are fixed at $\xi = 0.85$ and $\xi = 1.15$ across all distributions, and this is shown to provide nearly identical results in the Results section below.

Finally, we show the bias attributable to SLA when examining the distribution of capital (quantile) estimates based on 100,000 simulations, even when the true value of $\xi = 0.85$ , which is not particularly close to one. As described above, even under these circumstances, we can see the biasing effect of SLA's divergence distorting the results of the distribution of estimated quantiles (capital values).

**Table 1: Severity Distribution Parameter Values, Means, and Interpolation Points**

| Severity Distribution | Parameter1 | Parameter2 | Data collection threshold | Severity Distribution Mean | MISLA/ISLA Low Interpolation Point: $\xi$ = | MISLA/ISLA High Interpolation Point: $\xi$ = |
|---|---|---|---|---|---|---|
| BETAP | 5,000 | 1/0.99=1.010 | -- | 495,000 | 0.85 | 1.15 |
| FRCH | 1/0.99=1.010 | 5,000 | -- | 497,163 | 0.85 | 1.15 |
| GPD | 4,954.245 | 0.990 | -- | 495,425 | 0.8 | 1.2 |
| IGAM | 1/0.99=1.010 | 5,000 | -- | 495,000 | 0.85 | 1.15 |
| IPARA | 1/0.99=1.010 | 5,000 | -- | 500,050 | 0.85 | 1.15 |
| LOGG | 4.892 | 1/0.99=1.010 | -- | 6,069,948,738 | 0.9 | 1.2 |
| LOGL | 1/0.99=1.010 | 5,000 | -- | 495,081 | 0.85 | 1.15 |
| LOGN | 10.000 | 2.200 | -- | 247,707 | -- | -- |
| PARA | $\sqrt{1/0.99} = 1.005$ | 5,000 | -- | 495,041 | 0.925 | 1.075 |
| TGPD | 0.990 | 1,500 | 5,000 | 650,000 | 0.8 | 1.2 |
| TLOGG | 1.300 | 1/0.99=1.010 | 5,000 | 955,452 | 0.9 | 1.3 |
| TLOGN | 10.000 | 2.200 | 5,000 | 329,674 | -- | -- |

## Results

Summary

At both percentiles (99.9%tile and 99.97%tile), results of the empirical study show essentially identical results for MISLA and PE2 in terms of both accuracy and speed.[23] Both are between one and two orders of magnitude more accurate than SLA, and simulations of both show that the entire sampling distribution of SLA-based quantile (capital) estimates easily is distorted by its discontinuity even when the true value of the tail index is not close to a value of one. ISLA is slightly faster, but also slightly less accurate than both MISLA and PE2.

MISLA v. PE2

At both percentiles (99.9%tile and 99.97%tile), Tables 2 and 3, which use one billion simulated years of losses as the Monte Carlo benchmark, show that on average, PE2 is very slightly more accurate than MISLA, although excepting the Truncated LogGamma severity, MISLA is very slightly faster than PE2 (Table 4). However, both the speed and accuracy results for MISLA and PE2 are so close, especially in light of their standard deviations, that differences between them can be considered measurement error and thus, they are essentially identical. Since both are easily implemented, this leaves, as the only arguable advantage for MISLA, its consistency with the framework of those financial institutions already using SLA (but using it without an adjustment for its biasing discontinuity). Although determining the interpolation points for $\xi$ when using ISLA or MISLA is straightforward, and the specific values already listed in Table 1 above can be used, we did redo the analysis for MISLA using the fixed values for all severities of low $\xi = 0.85$ and high $\xi = 1.15$. The results remained virtually unchanged, and are shown in Table 5. These results notwithstanding, for severity distributions not covered in this paper, it is advisable to generate values of (4.a) and (4.c) as in Figure 1 to verify that the interpolation points used for the severities listed herein are appropriate for additional distributions.

MISLA vs. ISLA

For most severity distributions, the quantile (capital) approximation of ISLA is very close to that of MISLA (see Tables 2, 3, and 5). This indicates that an interpolation from (4.a) directly to (4.c), without anchoring on (4.b) in the middle of the 'bias zone,' usually is very good. The arguable exceptions here are the LogGamma, and to a lesser extent the Truncated LogGamma and Paralogistic distributions. For the LogGamma distribution (and its truncated version), the size of the 'bias zone' around $\xi = 1.0$ is much larger than those of the other severity distributions (see Figure 2 above), and hence, the approximation benefits from the use of (4.b) in its interpolation across the 'bias zone.' Because MISLA requires two interpolations and ISLA requires only one, the latter is slightly faster (see Table 4), although most would tradeoff the slight gain in speed for the slight gain in accuracy, mainly as a safety hedge as other distributions not covered herein may behave similarly to the LogGamma distribution.

MISLA vs. SLA

For the 99.9%tile, MISLA is almost two orders of magnitude more accurate than SLA, on average, even after discounting the extreme outlier of the LogGamma severity distribution (see Tables 2, 3, and 5). For the 99.97%tile, MISLA is well over an order of magnitude more accurate than SLA, again, even after discounting the

[23] CPU runtimes generally reflected real runtimes, and so the former are not reported herein.

**TABLE 2: Quantile Approximation of 99.9%Tile by Method by Severity Distribution**

| Severity | Monte Carlo* | SLA | ISLA | MISLA | PE2 |
|---|---|---|---|---|---|
| BETAP | 113,424,764 | 124,860,887 | 114,279,352 | 113,772,735 | 113,672,306 |
| FRCH | 114,035,893 | 125,388,379 | 114,770,013 | 114,252,764 | 114,154,474 |
| GPD | 114,020,697 | 125,444,154 | 114,505,122 | 114,280,595 | 114,198,969 |
| IGAM | 113,470,748 | 124,860,887 | 114,279,352 | 113,772,746 | 113,674,728 |
| IPARA | 115,121,597 | 126,587,847 | 115,809,679 | 115,337,411 | 115,240,687 |
| LOGG | 113,151,299 | 151,861,852,200 | 115,485,818 | 113,662,368 | 113,517,281 |
| LOGL | 113,981,342 | 125,334,105 | 114,721,232 | 114,142,225 | 114,099,303 |
| LOGN | 135,806,351 | 135,497,104 | 135,497,104 | 135,497,104 | 135,804,571 |
| PARA | 113,924,288 | 125,332,834 | 115,638,113 | 114,198,598 | 114,098,033 |
| TGPD | 148,688,257 | 163,440,790 | 149,225,234 | 148,792,144 | 148,800,580 |
| TLOGG | 141,746,694 | 164,352,571 | 143,161,736 | 141,896,625 | 141,905,369 |
| TLOGN | 159,022,573 | 158,563,746 | 158,563,746 | 158,563,746 | 158,998,022 |
| | %Difference | | | | |
| BETAP | 0% | 10.08% | 0.75% | 0.31% | 0.22% |
| FRCH | 0% | 9.96% | 0.64% | 0.19% | 0.10% |
| GPD | 0% | 10.02% | 0.42% | 0.23% | 0.16% |
| IGAM | 0% | 10.04% | 0.71% | 0.27% | 0.18% |
| IPARA | 0% | 9.96% | 0.60% | 0.19% | 0.10% |
| LOGG | 0% | 134111.32% | 2.06% | 0.45% | 0.32% |
| LOGL | 0% | 9.96% | 0.65% | 0.14% | 0.10% |
| LOGN | 0% | -0.23% | -0.23% | -0.23% | 0.00% |
| PARA | 0% | 10.01% | 1.50% | 0.24% | 0.15% |
| TGPD | 0% | 9.92% | 0.36% | 0.07% | 0.08% |
| TLOGG | 0% | 15.95% | 1.00% | 0.11% | 0.11% |
| TLOGN | 0% | -0.29% | -0.29% | -0.29% | -0.02% |
| | | | | | |
| | Mean (abs. value) | 9.62%** | 0.82% | 0.23% | 0.14% |
| | Std (abs. value) | 3.58%** | 0.51% | 0.10% | 0.08% |

*SAS® results, for which Monte Carlo = one billion simulated years.

** Excludes LogGamma.

**TABLE 3: Quantile Approximation of 99.97%Tile by Method by Severity Distribution**

| Severity | Monte Carlo* | SLA | ISLA | MISLA | PE2 |
|---|---|---|---|---|---|
| BETAP | 370,594,004 | 382,782,106 | 372,390,289 | 371,829,694 | 371,727,489 |
| FRCH | 373,006,679 | 384,459,098 | 374,032,853 | 373,459,961 | 373,345,016 |
| GPD | 373,415,315 | 384,737,281 | 373,977,687 | 373,711,360 | 373,626,769 |
| IGAM | 370,993,089 | 382,782,106 | 372,390,289 | 371,829,705 | 371,718,554 |
| IPARA | 376,685,880 | 388,249,170 | 377,653,300 | 377,136,251 | 377,037,936 |
| LOGG | 492,365,350 | 152,240,387,892 | 495,031,656 | 492,405,480 | 492,219,258 |
| LOGL | 372,956,126 | 384,404,851 | 373,979,279 | 373,335,764 | 373,304,628 |
| LOGN | 245,981,518 | 245,392,132 | 245,392,132 | 245,392,132 | 245,697,119 |
| PARA | 373,097,064 | 384,403,555 | 374,978,525 | 373,397,960 | 373,303,336 |
| TGPD | 486,136,360 | 501,017,735 | 487,054,930 | 486,549,685 | 486,552,834 |
| TLOGG | 472,024,860 | 494,756,339 | 473,914,599 | 472,480,435 | 472,479,487 |
| TLOGN | 284,218,910 | 283,848,178 | 283,848,178 | 283,848,178 | 284,260,414 |
| | %Difference | | | | |
| BETAP | 0% | 3.29% | 0.48% | 0.33% | 0.31% |
| FRCH | 0% | 3.07% | 0.28% | 0.12% | 0.09% |
| GPD | 0% | 3.03% | 0.15% | 0.08% | 0.06% |
| IGAM | 0% | 3.18% | 0.38% | 0.23% | 0.20% |
| IPARA | 0% | 3.07% | 0.26% | 0.12% | 0.09% |
| LOGG | 0% | 30820.21% | 0.54% | 0.01% | -0.03% |
| LOGL | 0% | 3.07% | 0.27% | 0.10% | 0.09% |
| LOGN | 0% | -0.24% | -0.24% | -0.24% | -0.12% |
| PARA | 0% | 3.03% | 0.50% | 0.08% | 0.06% |
| TGPD | 0% | 3.06% | 0.19% | 0.09% | 0.09% |
| TLOGG | 0% | 4.82% | 0.40% | 0.10% | 0.10% |
| TLOGN | 0% | -0.13% | -0.13% | -0.13% | 0.01% |
| | | | | | |
| | Mean (abs. value) | 2.97%** | 0.33% | 0.13% | 0.10% |
| | Std (abs. value) | 1.08%** | 0.14% | 0.08% | 0.08% |

*SAS® results, for which Monte Carlo = one billion simulated years.

** Excludes LogGamma.

**TABLE 4: SAS® Runtime (seconds) by Method by Severity Distribution**

| Severity | | ISLA | MISLA | PE2 |
|---|---|---|---|---|
| BETAP | | 0.0136 | 0.1400 | 0.2100 |
| FRCH | | 0.0131 | 0.0100 | 0.0400 |
| GPD | | 0.0123 | 0.0600 | 0.0100 |
| IGAM | | 0.0131 | 0.0900 | 0.0100 |
| IPARA | | 0.0127 | 0.0100 | 0.2000 |
| LOGG | | 0.0128 | 0.1700 | 0.1700 |
| LOGL | | 0.0126 | 0.0300 | 0.1500 |
| LOGN | | 0.0097 | 0.0097 | 0.0300 |
| PARA | | 0.0129 | 0.0400 | 0.1800 |
| TGPD | | 0.0146 | 0.0900 | 0.0100 |
| TLOGG | | 0.0151 | 0.8200 | 0.1700 |
| TLOGN | | 0.0108 | 0.0108 | 0.0300 |
| | | | | |
| | Median | 0.0129 | 0.0500 | 0.0950 |
| | Mean | 0.0128 | 0.1234 | 0.1008 |
| | Mean (sans TLOGG) | 0.0126 | 0.0600 | 0.0945 |

**TABLE 5: Accuracy of MISLA vs. MISLA with Fixed Interpolation Points, Across All Severity Distributions**

| Severity/ Platform | MISLA 99.9%tile | MISLA fixed pts 99.9%tile | Difference 99.9%tile | MISLA 99.97%tile | MISLA fixed pts 99.97%tile | Difference 99.97%tile |
|---|---|---|---|---|---|---|
| **SAS®** | | | | | | |
| GPD | 0.23% | 0.23% | 0.01% | 0.08% | 0.08% | 0.00% |
| LOGG | 0.45% | 0.48% | 0.03% | 0.01% | 0.02% | 0.01% |
| PARA | 0.24% | 0.20% | -0.04% | 0.08% | 0.07% | -0.01% |
| TGPD | 0.07% | 0.08% | 0.01% | 0.09% | 0.09% | 0.00% |
| TLOGG | 0.11% | 0.12% | 0.01% | 0.10% | 0.10% | 0.00% |

**TABLE 6: SLA vs. MISLA – One of the Larger Differences from 1000 Runs of 1000 Simulations (Each Run) for 10 Years of Losses (λ=25), GPD (θ = 4954.245; *ξ* = 0.85)**

| | (907 of 1000 simulations yielded $\hat{\xi} < 1.0$ ) | | | | | |
|---|---|---|---|---|---|---|
| | 99.9%tile | | | 99.97%tile | | |
| | SLA | MISLA | PE2 | SLA | MISLA | PE2 |
| Mean | 75,314,969 | 35,183,315 | 35,114,625 | 141,939,830 | 101,836,889 | 101,773,828 |
| Median | 26,214,717 | 26,200,076 | 26,107,661 | 69,691,786 | 69,660,250 | 69,587,902 |
| Minimum | 1,882,352 | 1,882,352 | 1,888,398 | 3,297,049 | 3,297,049 | 3,308,797 |
| Maximum | 35,709,569,403 | 156,084,006 | 155,988,732 | 35,921,777,662 | 514,084,824 | 513,988,495 |
| StdDev | 1,184,904,143 | 27,960,502 | 27,931,444 | 1,193,010,203 | 91,364,738 | 91,333,873 |
| Skewness | 30.09 | 1.25 | 1.25 | 29.84 | 1.39 | 1.39 |
| Kurtosis | 905.83 | 1.16 | 1.17 | 895.92 | 1.60 | 0.55 |

extreme SLA bias shown for the LogGamma severity distribution. In absolute terms, bias for $\xi = 0.99$ typically is on the order of magnitude of 10-30 million, although that of the LogGamma severity distribution is over 151 billion. There are two important, cautionary notes here. First, when the severity parameters, and thus tail index, are estimated based on samples of loss data, even when the true value of the tail index is not close to one its estimates can be, and oftentimes are, arbitrarily close to one. Consequently, for a given sample, SLA-based capital can be much more biased than is shown herein. And this is even more likely for constrained parameter estimations that restrict parameter values to those that yield only finite means (i.e. $\hat{\xi} < 1.0$ ), as these estimated values will be forced into the worst area of the 'bias zone' – very, very close to one – thus systematically and notably biasing capital upwards. This is ironic since the motivation for such constrained parameter estimation is to avoid unreasonably large capital estimates. Secondly, recall that the bias-inducing discontinuity of SLA is not an asymptotic result: it will hold regardless of sample size. So again, even when the true value of the tail index is not close to one, any simulation study easily can generate a few samples that bias the entire distribution of capital (quantile) estimates being generated. This result actually is shown in Table 6. One thousand simulations are run generating 1000 quantile (capital) approximations each for SLA and PE2 and MISLA based on the Generalized Pareto distribution with θ = 4954.245 and $\xi = 0.85$ (a tail index value not particularly close to one). This was then repeated 1000 times with 1000 different initial random number seeds. All of the 1000 mean values of the SLA-based quantiles were larger than those for the MISLA-based quantiles, and sometimes much larger: one of the largest differences of the 1000 groups of 1000 (but not the largest) is shown in Table 6. This particular group of 1000 happens to have a few samples of losses with estimated tail indices very close to one, yielding enormously biased capital estimates that distort the entire distribution of 1000. Given the frequency with which such simulations are run and are needed, the likelihood of material, systematic bias due to SLA's discontinuity is unacceptably high – hardly a freakishly rare occurrence, especially considering that it can occur with samples of any size. Of course, the story would be even more damaging for SLA when true tail index values are closer to one than $\xi = 0.85$ .[24]

Finally, it is worth noting that if this same exercise in Table 6 is performed for the LogGamma distribution, or similarly behaved distributions, the bias of the SLA-based quantile (capital) distributions would have been orders of magnitude larger.

[24] Maximum likelihood estimators were used in Table 6. The "true" quantile values for the 99.9 and 99.97%tiles, based on one billion years' worth of Monte Carlo simulations, are 32,496,320 and 89,383,135, respectively. Note that the respective means for both MISLA and PE2 are higher than these true values, indicating *estimation* bias in their quantile (capital) estimates, even though the *approximation* error of both is essentially zero. SLA's bias, which is predominantly approximation bias in this example, also includes this estimation bias, which will be present regardless of the accuracy of the approximation method used *and* regardless of the unbiasedness of parameter estimator used (including robust estimators). This is caused by Jensen's inequality – the fact that the quantile estimate is a convex function of the parameter estimate (see Jensen, 1906). This is thoroughly documented in Opdyke (2014), who identifies the three factors that determine the magnitude of this quantile bias: the heavier the severity distribution tail, the smaller the sample size, and the larger the quantile being estimated, the larger the quantile bias due to Jensen's inequality. Given that this setting requires the estimation of extreme quantiles of heavy-tailed loss distributions, Jensen's-induced quantile bias can be enormous, not only eclipsing possible bias from any other source, but also growing larger even than the quantile estimates themselves (see Opdyke, 2014, for empirical examples). Neither unbiased parameter estimation (robust or not), nor accurate approximation methods will address this: only methods developed specifically to mitigate Jensen's-induced quantile bias will do so (see Opdyke, 2014, for an example). Even though Table 6 examines a not-particularly-heavy-tailed severity distribution, empirically it still demonstrates that the three different sources of error described at the outset of this paper must be properly understood to appropriately deal with the separate effects of each: without this, we cannot hope to achieve our ultimate goal here, which is to obtain extreme quantile (capital) estimates that are reasonably accurate (unbiased), reasonably precise, and reasonably robust.

## Discussion and Conclusions

The objective of this paper was to develop a method to approximate extreme quantiles of heavy-tailed compound loss distributions that was simultaneously 1) very fast, 2) straightforward to implement, 3) consistent with widely used methods, and 4) very accurate. The Modified ISLA approximation (MISLA) satisfies all of these criteria. We compare its performance to the most viable competitor in the literature – the (second order) Perturbative Expansion approach of Hernandez et al. (2014) (PE2), which has been favorably compared to widely used competitors (such as Fast Fourier transform – see de Jong et al., 2016). We find essentially identical results between MISLA and PE2 in terms of both accuracy and speed. The only criterion by which MISLA arguably is preferable to PE2 is the third listed above, as it is simply a modified version of the Single Loss Approximation (SLA) that already is very widely used and incorporated into almost all AMA institutions' capital estimation frameworks.

Perhaps more importantly, even though SLA is the most widely used method of approximation to date (at least in operational risk modeling), it is clearly shown herein to possess a notable flaw in its implementation: namely, it has a discontinuity as values of the tail index of the severity distribution approach one from either side. This can systematically and very materially bias quantile (capital) estimates under a wide range of conditions, including 1) both finite and infinite means; 2) in simulation exercises even when true tail index values are not particularly close to one; and 3) regardless of sample size, as the discontinuity is a non-asymptotic result. The bias can be arbitrarily large since estimates of the tail index can be arbitrarily close to a value of one, so this method, as it stands, can provide very misleading quantile (capital) estimates in settings where the determination of very large amounts of capital depends directly on them. And constrained estimations that restrict severity parameters to values yielding only finite means do not solve this problem: in fact, they appear to exacerbate it. The improvement in accuracy of MISLA over SLA under the relatively conservative conditions tested herein is notable in both absolute and relative terms (one to two orders of magnitude), and in practice will be dramatic in many cases as the bias caused by SLA's discontinuity is unbounded.

Importantly, we note that aside from Opdyke (2014), no other paper has spelled out <u>why</u> the most widely used method in these settings often materially underperforms. Yet this understanding is absolutely necessary: absent this knowledge, we cannot know definitively where it will underperform, let alone know how to fix it so that a successor – in this case, MISLA – can be as good or better than its competitors.

We conclude that both MISLA and PE2 are easily implemented alternatives to SLA, and that based on the evidence provided herein, both methods can be considered among the best to use for these purposes in these settings: namely, for the fast and accurate approximation of extreme quantiles of heavy-tailed compound loss distributions, which are commonly found in operational risk, insurance, and catastrophic loss models.

# Appendix

**Beta-prime Distribution:** This distribution is also known as the inverted beta distribution or the beta distribution of the second kind (see Johnson et al., 1995).

PDF: $f\left(x;\alpha,\beta\right)=\dfrac{x^{\alpha-1}}{\mathrm{B}\left(\alpha,\beta\right)\left(1+x\right)^{\alpha+\beta}}$, $0<x<\infty,\ 0<\alpha<\infty,\ 0<\beta<\infty,$

where $\mathrm{B}\left(\alpha,\beta\right)=\int_0^1 x^{\alpha-1}\left(1-x\right)^{\beta-1}dx$ is the Beta function.

CDF: $F\left(x;\alpha,\beta\right)=I_{(x/x+1)}\left(\alpha,\beta\right)=B\left(\left[x/\left(x+1\right)\right],\alpha,\beta\right)\big/B\left(\alpha,\beta\right),\ 0<x<\infty,\ 0<\alpha<\infty,\ 0<\beta<\infty$ where

$$B\left(z,\alpha,\beta\right)=\int_0^z t^{\alpha-1}\left(1-t\right)^{\beta-1}dt$$

Alternately, $F\left(x;\alpha,\beta\right)=F_{Beta}\left(\dfrac{x}{1+x};\alpha,\beta\right)$, where $F_{Beta}\left(\cdot;\alpha,\beta\right)$ is the CDF of the Beta distribution.

Mean: $\mathrm{E}\left(X\right)=\dfrac{\alpha}{\beta-1},\ \beta>1$ Tail Index: $\xi=\dfrac{1}{\beta}$

**Frechet Distribution:** This distribution is also known as the type II extreme value distribution, log-Gompertz or Inverse Weibull distribution (see Kleiber and Kotz, 2003).

PDF: $f(x;a,b)=\frac{a}{b}\left(\frac{b}{x}\right)^{a+1} e^{-\left(\frac{b}{x}\right)^a},\ 0<x<\infty,\ 0<a<\infty,\ 0<b<\infty.$

CDF: $F(x;a,b)=e^{-\left(\frac{b}{x}\right)^a},\ 0<x<\infty,\ 0<a<\infty,\ 0<b<\infty$

Mean: $\mathrm{E}(X)=b\Gamma\left(1-\frac{1}{a}\right),\ a>1,$ where $\Gamma(s)=\int_0^\infty t^{s-1}e^{-t}dt$ is the Gamma function Tail Index: $\xi=\frac{1}{a}$

**Generalized Pareto Distribution:**

PDF: $f(x;\xi,\vartheta)=\frac{1}{\vartheta}\left(1+\xi\frac{x}{\vartheta}\right)^{-\frac{1}{\xi}-1},\ 0\le x<\infty,\ 0\le\xi<\infty,\ 0<\vartheta<\infty.$

Note that we have restricted parameter range for $\xi$ from $\mathbb{R}$ to $\mathbb{R}_0^+$ to follow common industry practice.

CDF: $F(x;\xi,\vartheta)=1-\left(1+\xi\frac{x}{\vartheta}\right)^{-\frac{1}{\xi}},\ 0\le x<\infty,\ 0\le\xi<\infty,\ 0<\vartheta<\infty$

Mean: $\mathrm{E}(X)=\frac{\vartheta}{1-\xi},\ 0\le\xi<1$ Tail Index: $\xi=\xi$

**Left Truncated Generalized Pareto Distribution:**

PDF: $g(x;\xi,\vartheta)=\frac{f(x;\xi,\vartheta)}{1-F(H;\xi,\vartheta)},\ H\le x<\infty,\ 0\le\xi<\infty,\ 0<\vartheta<\infty,$

where $f$ and $F$ are the PDF and CDF of the Generalized Pareto distribution.

Equivalently, $g(x;\xi,\vartheta)=(H\xi+\vartheta)^{\frac{1}{\xi}}(x\xi+\vartheta)^{-\frac{1}{\xi}-1}$.

CDF: $G(x;\xi,\vartheta)=\frac{F(x;\xi,\vartheta)-F(H;\xi,\vartheta)}{1-F(H;\xi,\vartheta)},\ H\le x<\infty,\ 0\le\xi<\infty,\ 0<\vartheta<\infty,$

or, equivalently, $G(x;\xi,\vartheta)=1-(H\xi+\vartheta)^{\frac{1}{\xi}}(x\xi+\vartheta)^{-\frac{1}{\xi}}$.

Mean: $\mathrm{E}(X)=\frac{H+\vartheta}{1-\xi},\ 0\le\xi<1$ Tail Index: $\xi=\xi$

**Inverse Gamma Distribution:** This distribution is also known as the Vinci distribution (see Kleiber and Kotz, 2003).

PDF: $f(x;a,b)=\frac{b^a}{\Gamma(a)}x^{-a-1}e^{-\left(\frac{b}{x}\right)},\ 0<x<\infty,\ 0<a<\infty,\ 0<b<\infty,\ \Gamma(s)=\int_0^\infty t^{s-1}e^{-t}dt$, the Gamma function.

CDF: $F(x;a,b)=\Gamma(a,b/x)/\Gamma(a),\ 0<x<\infty,\ 0<a<\infty,\ 0<b<\infty,$ where $\Gamma(q,p)=\int_p^\infty t^{q-1}e^{-t}dt$

Alternately, $F(x;a,b) = 1 - J\left(\frac{1}{x};\alpha,\frac{1}{\beta}\right)$ where $J\left(\cdot;\alpha,\frac{1}{\beta}\right)$ is the CDF of the Gamma distribution.

<u>Mean</u>: $\mathrm{E}(X) = \frac{b}{a-1},\ a > 1$ <u>Tail Index</u>: $\xi = \frac{1}{a}$

**Inverse Paralogistic Distribution:** See Kleiber and Kotz, 2003 (mean is for $p = a$).

<u>PDF</u>: $f(x;a,b) = \frac{a^2 x^{a^2-1}}{b^{a^2}\left(1+\left(\frac{x}{b}\right)^a\right)^{a+1}},\ 0 < x < \infty,\ 0 < a < \infty,\ 0 < b < \infty.$

<u>CDF</u>: $F(x;a,b) = \frac{1}{\left(1+\left(\frac{b}{x}\right)^a\right)^a},\ 0 < x < \infty,\ 0 < a < \infty,\ 0 < b < \infty$

<u>Mean</u>: $\mathrm{E}(X) = \frac{b\left(\Gamma\left(1-\frac{1}{a}\right)\Gamma\left(a+\frac{1}{a}\right)\right)}{\Gamma(a)},\ a > 1,\ \Gamma(s) = \int_0^\infty t^{s-1}e^{-t}dt$, the Gamma function <u>Tail Index</u>: $\xi = \frac{1}{a}$

**LogGamma Distribution:** See Kleiber and Kotz, 2003.

<u>PDF</u>: $f(x;\alpha,\beta) = \frac{1}{x^{\beta+1}\Gamma(\alpha)}(\log(x))^{\alpha-1}\beta^{\alpha},\ 1 \le x < \infty,\ 0 < \alpha < \infty,\ 0 < \beta < \infty,\ \Gamma(s) = \int_0^{+\infty} t^{s-1}e^{-t}dt$

<u>CDF</u>: $F(x;\alpha,\beta) = \int_1^x f(t;\alpha,\beta)dt,\ 1 \le x < \infty,\ 0 < \alpha < \infty,\ 0 < \beta < \infty.$

Alternately, $F(x;\alpha,\beta) = J\left(\log(x);\alpha,\frac{1}{\beta}\right)$, where $J\left(\cdot;\alpha,\frac{1}{\beta}\right)$ is the CDF of the Gamma distribution.

<u>Mean</u>: $\mathrm{E}(X) = \left(\frac{\beta}{\beta-1}\right)^{\alpha},\ \beta > 1$ <u>Tail Index</u>: $\xi = \frac{1}{\beta}$

**Left Truncated Loggamma Distribution:** See Opdyke (2014).

<u>PDF</u>: $g(x;\alpha,\beta) = \frac{f(x;\alpha,\beta)}{1-F(H;\alpha,\beta)},\ \mathrm{H} \le x < \infty,\ 0 < \alpha < \infty,\ 0 < \beta < \infty,$ where $f$ and $F$ are the PDF and CDF of the LogGamma distribution.

<u>CDF</u>: $G(x;\alpha,\beta) = \frac{F(x;\alpha,\beta) - F(H;\alpha,\beta)}{1-F(H;\alpha,\beta)},\ \mathrm{H} \le x < \infty,\ 0 < \alpha < \infty,\ 0 < \beta < \infty.$

<u>Mean</u>: $\mathrm{E}(X) = \left(\frac{\beta}{\beta-1}\right)^{\alpha}\frac{1-J(\log(H)(\beta-1);\alpha,1)}{1-F(H;\alpha,\beta)},\ \beta > 1$,

where $J(\bullet)$ = CDF of the Gamma distribution. Alternately, $\mathrm{E}(X)=\left(\frac{\beta}{\beta-1}\right)^{\alpha}\frac{1-F(H;\alpha,\beta-1)}{1-F(H;\alpha,\beta)},\ \beta>1$

Tail Index: $\xi=\frac{1}{\beta}$

**LogLogistic Distribution:**

PDF: $f(x;a,b)=\frac{ax^{a-1}}{b^{a}\left(1+\left(\frac{x}{b}\right)^{a}\right)^{2}},\ 0<x<\infty,\ 0<a<\infty,\ 0<b<\infty.$

CDF: $F(x;a,b)=\frac{1}{\left(1+\left(\frac{b}{x}\right)^{a}\right)},0<x<\infty,\ 0<a<\infty,\ 0<b<\infty$

Mean: $\mathrm{E}(X)=\frac{b\Gamma\left(1+\frac{1}{a}\right)\Gamma\left(1-\frac{1}{a}\right)}{\Gamma(1)},\ a>1,$ which simplifies to $\mathrm{E}(X)=\frac{b\pi}{a\sin\left(\frac{\pi}{a}\right)},\ a>1$

where $\Gamma(s)=\int_{0}^{\infty}t^{s-1}e^{-t}dt$ is the Gamma function Tail Index: $\xi=\frac{1}{a}$

**LogNormal Distribution:**

PDF: $f(x;\mu,\sigma)=\frac{1}{\sigma x\sqrt{2\pi}}e^{-\frac{1}{2}\left(\frac{\ln(x)-\mu}{\sigma}\right)^{2}},\ 0<x<\infty,\ -\infty<\mu<\infty,\ 0<\sigma<\infty.$

CDF: $F(x;\mu,\sigma)=\frac{1}{2}\left(1+\mathrm{erf}\left(\frac{\ln(x)-\mu}{\sqrt{2\sigma^{2}}}\right)\right),$ where $\mathrm{erf}(x)=\frac{2}{\sqrt{\pi}}\int_{0}^{x}e^{-t^{2}}dt$ is the error function.

Mean: $\mathrm{E}(X)=e^{\mu+\frac{\sigma^{2}}{2}}$ Tail Index: Not Applicable

**Left Truncated LogNormal Distribution:**

PDF: $g(x;\mu,\sigma)=\frac{f(x;\mu,\sigma)}{1-F(H;\mu,\sigma)},\ H<x<\infty,\ -\infty<\mu<\infty,\ 0<\sigma<\infty,$

where $f$ and $F$ are the pdf and CDF of the Lognormal distribution.

CDF: $G(x;\mu,\sigma)=\frac{F(x;\mu,\sigma)-F(H;\mu,\sigma)}{1-F(H;\mu,\sigma)},\ H<x<\infty,\ -\infty<\mu<\infty,\ 0<\sigma<\infty.$

Mean: $\mathrm{E}\left(X\right)=\Phi\left(\frac{\mu+\sigma^2-\log\left(H\right)}{\sigma}\right)\frac{e^{\mu+\frac{\sigma^2}{2}}}{1-F\left(H;\mu,\sigma\right)}$ Tail Index: Not Applicable

**Paralogistic Distribution:**

PDF: $f\left(x;a,b\right)=\frac{a^2x^{a-1}}{b^a\left(1+\left(\frac{x}{b}\right)^a\right)^{1+a}},\ 0<x<\infty,\ 0<a<\infty,\ 0<b<\infty.$

CDF: $F\left(x;a,b\right)=1-\frac{1}{\left(1+\left(\frac{x}{b}\right)^a\right)^a},\ 0<x<\infty,\ 0<a<\infty,\ 0<b<\infty$

Mean: $\mathrm{E}\left(X\right)=\frac{b\left(\Gamma\left(1+\frac{1}{a}\right)\Gamma\left(a-\frac{1}{a}\right)\right)}{\Gamma\left(a\right)},\ a>1,$ where $\Gamma\left(s\right)=\int_0^\infty t^{s-1}e^{-t}dt$ Tail Index: $\xi=\frac{1}{a^2}$